\documentclass[preprint,12pt,authoryear]{elsarticle}
\usepackage{amsmath,amsfonts,amssymb,amsthm}
\usepackage{color}
\usepackage{pbox}
\usepackage{lineno}
\usepackage{caption}
\usepackage{subfig}
\usepackage{array, longtable, tabularx}
\usepackage{adjustbox}
\usepackage{pdflscape}
\usepackage{hyperref}
\modulolinenumbers[5]
\let\oldtabular\tabular 
\renewcommand{\tabular}{\scriptsize\oldtabular}
\def\arcsec{\hbox{$^{\prime\prime}$}}

\begin{document}

\begin{frontmatter}
\title{Benchmarking information carriers}
\author{Michael Hippke\corref{label3}}
\ead{michael@hippke.org}
\address{Sonneberg Observatory, Sternwartestr. 32, 96515 Sonneberg, Germany}

\begin{abstract}
The search for extraterrestrial communication has mainly focused on microwave photons since the 1950s. We compare other high speed information carriers to photons, such as electrons, protons, and neutrinos, gravitational waves, inscribed matter, and artificial megastructures such as occulters. The performance card includes the speed of exchange, information per energy and machine sizes, lensing performance, cost, and complexity. In fast point-to-point communications, photons are superior to other carriers by orders of magnitude. Sending probes with inscribed matter requires less energy, but has higher latency. For isotropic beacons with low data rates, our current technological level is insufficient to determine the best choice. We discuss cases where our initial assumptions do not apply, and describe the required properties of hypothetical particles to win over photons.
\end{abstract}
\end{frontmatter}

\section{Introduction}
Information carriers for interstellar communication can be electromagnetic waves (photons), other particles such as neutrinos or protons, gravitational waves and inscribed matter (probes). The search for extraterrestrial intelligence (SETI), which is in reality mainly a search for communications, focuses heavily on photons, particularly at microwave and optical frequencies \citep{1985IAUS..112..271T,2001ARA&A..39..511T,2010SPIE.7819E..02T}. For photons, the search space is 9-dimensional: three of space, two of each polarization, intensity, modulation, frequency, and time. For other carriers, the search space has not been described in much detail yet. For example, Neutrinos come without polarization, but as particle/antiparticle and have three flavors (electron, muon, tau). A consolidated review of advantages and disadvantages for all carriers is missing in the literature accessible to us. In this paper, we will review all known possibilities for communication and benchmark them against photons. The academic literature indicates a rising and significant dissatisfaction with the orthodox SETI approach, which focuses almost exclusively on radio observations \citep{2011JBIS...64..156B,2014ApJ...792...27W,2014ApJ...792...26W}. Therefore, it is important to consider all alternative communication methods.

While we do not know the motivations and goals of other civilizations, if they exist, we here assume the universal value of exchanging more information rather than less. Other preferences which appear natural to us are the speed of exchange (a preference to obtain information earlier than later), the amount of energy required (obtaining as much information as possible for a finite amount of resources), and the ease of use (although this is difficult to measure). These aspects were already identified in the Cyclops Report, often dubbed the ``SETI bible'' \citep{1971asee.nasa.....O}. We will relax these constraints in section~\ref{irrational_aliens}.

In this work, we are largely agnostic to ever-changing technological cost functions, and focus on the physical optimum. We systematically compare all known communication methods with regards to their energy efficiency and data rates over pc to kpc distances. We benchmark them against photons, which have a capacity (in units of bits per energy) of \citep{2017arXiv171105761H}

\begin{equation}
\label{eq_photon}
C_{\gamma} = 
\left(\frac{d}{1\,{\rm pc}} \right)^{-2} 
\left(\frac{\lambda}{1\,{\rm nm}} \right)^{-1} 
\left(\frac{D}{1\,{\rm m}} \right)^{4}
\,\,(\rm {bits\,J^{-1}})
\end{equation}

where $d$ is the distance, 
$\lambda$ is the photon wavelength,
$D=D_{\rm t}=D_{\rm r}$ are equal-sized transmitter and receiver apertures,
and a conservative information efficiency of one bit per photon is assumed \citep{2017arXiv171205682H,2018arXiv180106218H}. This estimate is valid to within 10\,\% neglecting losses and noise.

Artificial occulters and other physical objects are treated in section~\ref{occulter},
followed by charged (section~\ref{charged_particles}),
short-lifetime (section~\ref{short_particles}) and
massive (section~\ref{massive_particles}) particles;
gravitational waves (section~\ref{gravitational_waves}) and 
neutrinos (section~\ref{neutrinos}).

\section{Artificial occulters, inscribed matter, and other physical objects}
\label{occulter}
For an information exchange, one can distinguish between scenarios where the transmitter controls only the energy output (category I, e.g., modulating the energy output of a Cepheid variable star), the directivity (category II, e.g., an occulting screen in front of a star), or both (category III, e.g., a laser).

\subsection{Artificial occulters}
\label{arti}
Exotic examples for category II communications are artificial megastructures occulting a star, such as the geometrical objects described by \citet{2005ApJ...627..534A}. The advantage is that the energy source is already available, broadband, and powerful. Directivity is defined by the orbits of the structures, and limited by the fixed size of the star, which produces a wide (few degree for planet-sized structures at au distances) cone of visibility targeting $\mathcal{O}(10^6)$ stars within a few hundred pc. We may assume that curious civilizations would monitor nearby stars for (natural) exoplanet transits, and thus detect the non-spherical transiting structures. The obvious disadvantages are the need to build space-based megastructures with sizes $10^3 \dots 10^5$\,km, and the low information content. Encoding schemes could use transit timing and -duration variations and shape modifications, but are limited to a few bits per transit by the small number of these modes and stellar noise.

Megaengineering constructions may be intended for many purposes, such as shielding swarms \citep{2016AcAau.129..438C}, and their visibility as communication beacons might be, deliberate or not, a side effect.

A modification of the transit idea, falling into category I, is the use of artificial light sent from Earth during a ``planet Earth transit'' which modifies the observed spectral transit signature of our planet \citep{2016MNRAS.459.1233K}. 

Other large structures described in the literature are (partial) \citet{1960Sci...131.1667D} spheres or some of their modifications such as \citet{1987brig.iafcR....S} thrusters,  spherical arc mirrors to use the impulse from a star's radiation pressure \citep{2013JBIS...66..144F}, or starshades to reduce the irradiation from evolving stars \citep{2017MNRAS.469.4455G}. Such objects are not primarily built for communication and thus neglected here.

Similarly, other bright natural sources might be used as beacons, such as pulsars \citep{2003IJAsB...2..231E} or supernovae \citep{1994Ap&SS.214..209L}. Very advanced civilizations may modify the pulsation cycles of Cepheid variable stars as beacons to transmit all-call information throughout the galaxy and beyond \citep[category I,][]{2012ConPh..53..113L,2015ApJ...798...42H}.

Some sources might appear as laser or microwave communication (category III), but could in fact be artificial directed energy to push light sails; these would appear as transient events with flux densities of Jy and durations of tens of seconds at 100 pc \citep{2015ApJ...811L..20G,2017ApJ...837L..23L}.

\subsection{Inscribed matter}
Sending a physical artifact can be the most energy-efficient choice \citep{2004Natur.431...47R}, because it can be done at almost arbitrarily low velocities, and thus low energies. Also, an artifact can arrive at the destination in total, in contrast to a beam which is wider than the receiver in all realistic cases, so that most energy is lost. The obvious disadvantage is the large communication delay, e.g. sending a probe at a (relatively fast) $0.01\,c$ takes 438 years to the nearest star. We argue that inscribed matter on board of exploration probes could make sense, e.g. holding a galactic library of knowledge to be discovered by others, including a tutorial for communication. Then, the first contact would ease the need for fast communication, as reading the library could keep the impatient Aliens busy while waiting for the first ``answer'' from the stars. An in-depth analysis is given in \citet{2017arXiv171210262H}, which finds the capacity of the inscribed matter channel as

\begin{equation}
\label{eq_matter}
C_{\rm matter} = \eta\, S\, L^{-1}\, v^{-2} \,\,(\rm {bits\,J^{-1}}).
\end{equation}

where $\eta$ is the propulsion efficiency factor for launch and deceleration, $S_{\rm base}$ is the information density per unit mass, and $L$ is the relativistic Lorentz factor. For $S_{\rm base}=10^{23}$\,bits\,g$^{-1}$ and $v=0.1\,c$, we have an energy efficiency of $\approx10^{11}$\,bits per Joule. A photon channel requires large apertures to match this efficiency due to diffraction losses. For example, $d=1.3\,$pc requires $D = 1\,$km apertures at X-ray energy (1000\,km at $\lambda=1\,\mu$m). For a constant wavelength, apertures must increase with distance as $D \propto d^{1/2}$, making probes more attractive for large amounts of data over large distances. For low probe speeds ($v \ll 0.01\,$c), equivalent aperture sizes are implausibly high. Inscribed matter is more energy efficient by many orders of magnitude in this regime of slow probes.

Apart from a communication link between two distant species, an inscribed matter artefact can be placed in a stellar system, to be found at a later date \citep{2012AcAau..78..121C}. This scenario was popularized by Arthur C. Clarke's moon monolith \citep{Clarke1953}. Searches for artefacts inside our solar system have been suggested \citep{1960Natur.186..670B,1995ASPC...74..425P,2004IAUS..213..487T,2012AcAau..72...15H}, as well as for starships \citep{Martin1980}. Possible locations include in geocentric, selenocentric, Earth-Moon libration, Earth-Moon halo orbits \citep{1980Icar...42..442F,1983Icar...53..453V,1983Icar...55..337F}, the Kuiper belt \citep{2012AsBio..12..290L}, and even ``footprints of alien technology on Earth'' \citep{2012AcAau..73..250D}.

\subsection{Assessment}
Leveraging available power sources as all-sky beacons appears attractive, but directivity requires building megastructures (e.g., a large Fresnel lens, occulters etc.). Such is physically possible, but the energy needed in the build process might be better spent in actual communication, e.g. using lasers. Structures such as small (planet-sized) occulters can not leverage the isotropic radiation; they only harvest a small fraction of the starlight into modulated beacons.

Throughout the paper we have made the assumption that the main purpose of a machine is communication, and all related costs (Table~\ref{table1}) are attributed to it. This may be false, as many buildings on Earth, such as skyscapers, serve double duty as office buildings and antenna carriers. Determining the actual communication costs, without the commercial part, is not trivial, and will be relevant for future SETI efforts with maturing technology \citep{Harrison2009}.

On grand scale, even the cosmic microwave background might provide an opportunity to send a message to all occupants in the universe \citep{2006MPLA...21.1495H}, although the information content might be limited to less than 1000 (unchangeable) bits \citep{2005physics..11135S}.

We concede that it is crucial to keep an open mind for unknown phenomena \citep{2017arXiv170805318W}, perhaps serving as beacons \citep{2016ApJ...816...17W}. However we can not see an attractive, yet superior, communication option in the schemes described in the literature when it comes to targeted communication with high data rates.

\begin{landscape}
\fontsize{10}{10}\selectfont
\centering
\begin{longtable}{lcccccccccc}
\caption{Carrier performance summary \label{tab:particle_overview}}
\label{table1}\\
\hline
Carrier & MeV/c$^2$ & Lifetime (s) & Velocity ($c$)  & Charged & \pbox{20cm}{Beam\\angle\,($\arcsec$)} & Extinction & Lensing & Difficulty & \pbox{20cm}{Initial\\cost} & \pbox{20cm}{Operational\\ cost} \\
\hline
\endhead
\endfoot

\endlastfoot
Photon   & 0        & stable                      & 1  & no & $10^{-4}$ & 0.001  & good & low & medium & medium \\
Neutrino & $\approx 0.001$ & oscillations         & $\lessapprox 1$ & no & $\textcolor{red}{10^6}$ & $\approx 0$ & very good & medium & \textcolor{red}{high} & medium \\
Electron & 0.51     & $>10^{36}$                  & \textcolor{red}{0.0630} &  \textcolor{red}{yes} & \textcolor{red}{$10^6$}  & medium & medium& \textcolor{red}{$\approx 1$}& medium & medium \\
Proton   & 938.27   & $>10^{51}$                  & \textcolor{red}{0.0015} &  \textcolor{red}{yes} & \textcolor{red}{$10^6$} & \textcolor{red}{$\approx 1$}& very good & low & medium & medium\\
Neutron  & 939.56   & \textcolor{red}{$881.5$}    & \textcolor{red}{0.0015} &  no & \textcolor{red}{$10^6$} & \textcolor{red}{$\approx 1$}& very good & low & medium & medium\\
Muon     & 105.66   & \textcolor{red}{$10^{-6}$}  & \textcolor{red}{0.0044} &  \textcolor{red}{yes} & \textcolor{red}{$10^6$} & \textcolor{red}{$\approx 1$}& very good & low& \textcolor{red}{high} & medium\\
Tau      & 1776.82  & \textcolor{red}{$10^{-13}$} & \textcolor{red}{0.0011} &  \textcolor{red}{yes} & \textcolor{red}{$10^6$} & \textcolor{red}{$\approx 1$}& very good & low& \textcolor{red}{high} & medium\\
Higgs & $1.25\times10^5$ & \textcolor{red}{$10^{-22}$} & \textcolor{red}{$10^{-6}$} & no & \textcolor{red}{$10^6$} & $\approx 0$ & very good & \textcolor{red}{high} & \textcolor{red}{high} & \textcolor{red}{high}\\

\hline
Inscribed matter & $>0$ & $>10^{13}$ & $0 \dots 1$ & no & $\approx 0$ & $\approx 0$ & no & low&\textcolor{red}{high} & zero\\
Occulter & $>0$ & $>10^{13}$ & 1 & possible & \textcolor{red}{$10^3$} & 0.001 & unlikely & low &\textcolor{red}{high} & very low\\
Gravitational wave & 0 & stable & 1 & no & \textcolor{red}{$4\pi$} & 0 & low &\textcolor{red}{high}&\textcolor{red}{high}&\textcolor{red}{high}\\
\hline
Axion & $\approx10^{-9}$ & stable & $\lessapprox 1$ & no & $10^{-4}$ & $\approx$ 0 & good & \textcolor{red}{high} & unclear & unclear  \\
Tachyon & unclear & stable & $>1$ & no & unclear & unclear & unclear & \textcolor{red}{high} & unclear & unclear \\
\hline

\end{longtable}
All values for $E={\rm keV}$, transmitter aperture 1\,m, distance 1\,pc. Masses and lifetimes from \citet{Mohr2006}. Text in \textcolor{red}{red color} indicates problematic properties.
\end{landscape}

\section{Charged and massive particles}
This section covers all particles except neutrinos and photons.

\subsection{Charged particles}
\label{charged_particles}
Particles with an electric charge, such as an electron (positron), proton, muon or tau, are deflected by magnetic fields and absorbed by the interstellar medium. It is expected that these particles lose any correlation with their original direction as they traverse through interstellar magnetic fields. In reality, small-scale anisotropies have been observed \citep{2017arXiv170803005T} which are believed to originate from structures in the heliomagnetic
field, turbulence in galactic magnetic fields, and non-diffusive propagation \citep{2017PrPNP..94..184A}. Despite these anisotropies, it is highly unlikely that even perfect knowledge of the interstellar magnetic fields would allow for communication channels that are stable over useful timescales. Thus, we argue that charged particles are not usable for interstellar communication based on known physics.

\subsection{Particles with short lifetimes}
\label{short_particles}
Some particles like the muon or tau decay on very short timescales and are thus unsuitable. For example, muons with a lifetime of $10^{-6}$\,s (Table~\ref{table1}) decay within seconds even for high energies (PeV, Lorentz factor $\gamma \approx 10^6$). For a limited range to the nearest stars, a particle lifetime of a few years is required. For uncharged particles, this is only possible with Neutrons, as their half-life at $\mathcal{O}(100)$\,TeV, $\gamma \approx 10^5$ is $\approx 2.8$\,yrs. Such neutrons are easy to detect using atmospheric Cherenkov detectors \citep{2006ApJ...636..777A} and have low cosmic ray background (i.e. noise). Using muons (taus) for interstellar communication with year-long travel times would require extreme particle energies of $\mathcal{O}(10^{4})$\,EeV ($\mathcal{O}(10^{12})$\,EeV for taus), making these choices extremely energy inefficient. These energies are above Planck energy $E_{\rm P}=\sqrt{\hbar c^5/G}$ and thus physically implausible, because such high energy particles can not be produced, and would directly collapse into a black hole \citep{1995hep.ph...10364C}.

\subsection{Massive particles}
\label{massive_particles}
Some particles are heavy which makes it costly to accelerate them (proton, neutron, muon, tau). The Higgs boson has the highest mass of all known elementary particles ($125\,$GeV/c$^2$) and decays after $10^{-22}$\,s. Interstellar communication would thus require (unphysical) particle energies of $\mathcal{O}(10^{22})$\,EeV; substituting one Higgs would allow for sending $\mathcal{O}(10^{36})$\,keV photons instead, making this choice irrational.

\subsection{Hypothetical particles}
\label{hypo}
It is sometimes argued in the hallways of astronomy departments that we ``just have to tune into the right band'' and -- voil\`{a} -- will be connected to the galactic gossip channel with the latest and greatest news about the Princess of Betelgeuse's\footnote{The M2 Iab red supergiant Betelgeuse (\textit{$\alpha\,$Orionis}) is expected to explode in a supernova in $<10^6$ years \citep{2013EAS....60...17M,2016ApJ...819....7D}. At a distance of $222 \substack{+48\\-38}\,$pc \citep{2017AJ....154...11H} it might well have exploded as of now, without us knowing yet.} tragic death (R. H., priv. comm.). 
The problem with this argument is that isotropic radiation, filling the entire galaxy with information-carrying particles of any kind, would require a prohibitively large amount of energy due to the minimum requirement of energy per bit of information, $kT\ln2$ \citep{1998RSPSA.454..305L}. When it can not be isotropic, it must instead be targeted. So far, we have not detected any emission in the most trivial bands such as microwaves or optical. But could it be some yet unknown particle?

A candidate carrier for such a novel communication channel is the axion, a hypothetical particle postulated by \citet{1977PhRvD..16.1791P,1977PhRvL..38.1440P} to resolve the strong CP problem in quantum chromodynamics. The axion, if it exists, would have a low mass of order $10^{-5}$\,eV, which is negligible compared to the energy used for communication purposes at $\mu$m wavelength and below (eV energy and above). Axions, like neutrinos, do not interact much with matter, which makes extinction negligible, but detection difficult. The advantage with axions is that the signals at the input and output would be electromagnetic waves, so that existing modulation technology could be used. Despite the lack of strong interactions with matter, axions might be detected via their coupling to the electromagnetic field \citep{1985PhRvD..32.2988S}. A shielded, cooled, sharp resonator tuned to the right frequency could work as a receiver; given that the correct frequency is known to great (order Hz) accuracy \citep{2007PhRvD..76k1701S}. If the frequency is unknown, the receiver would need to sample the frequency space, or fall back to some magic frequency, which we can not yet guess given the unknown nature of axions.

Among the more ``radical'' speculations are objects such as stable remnants of the process of Hawking evaporation of black holes \citep{1993PhRvD..47..540B,2005JHEP...10..053K}. If such objects exist, they might be excellent information repositories, with the highest possible information density, thus presenting a microscopic analogue of the inscribed matter.

The advantage of the axion over the photon is solely in its lower extinction. This is a small advantage, given that there are many photon wavelengths with low ($<1$\%) extinction over pc, and even kpc, distances \citep{2017arXiv170603795H}. The added complexity in axion transmission and reception seems hardly worth the effort, unless one wants to deliberately set an entrance barrier for the primitive civilizations.

We note that there is a zoo of other hypothetical particles which could be discussed in a similar way, e.g. ``hidden'' photons \citep{2009EL.....8710010J}, or the Neutralino \citep{2010ARA&A..48..495F}. The discussion on such particles will be carried out in section~\ref{we_know_nothing}.

\subsection{Assessment}
\label{acc}
The only realistic candidate from this category is the neutron. The major issue for communication with neutrons is energy efficiency. Known physics only offers the focusing of such particle beams in accelerators. The beam width produced by an accelerator scales as \citep{2009JInst...4T5001I}

\begin{equation}
\theta_{\rm beam} = \frac{1}{\gamma} \approx \frac{10^{-4}}{E_{\mu} {\rm [TeV]}}
\end{equation}

where $\gamma$ is the relativistic boost factor of a muon, and $E_{\mu}$ is its energy. The beam angle at GeV (TeV, PeV) energies is $6^{\circ}$ (21\,arcsec, 21\,mas). For comparison, the opening angle of diffraction-limited optics is $\theta_{\rm optics} = 1.22 \lambda / D_{\rm t}$. For $D_{\rm t}=1$\,m, $\theta_{\rm beam} = \theta_{\rm optics} $ at $\lambda=82$\,nm (15 eV), a difference of $7\times10^{10}$ in energy for the same beam width. In other words, focusing particles into a beam requires $7\times10^{10}$ more energy in a particle accelerator (using TeV particles) compared to a meter-sized mirror (with 15\,eV photons). 

Practical examples for beam angles can be found from keV \citep{2012NaPho...6..308T,2015PhPl...22b3106T}, MeV \citep{PhysRevLett.110.155003,2014OptL...39.4132L,2017arXiv170508637S} to GeV energies, for a variety of particles, including neutrinos \citep{2008AcPPB..39.2943S}. 

An additional limit for high energy particles arises near the Greisen-Zatsepin-Kuzmin limit \citep{1966PhRvL..16..748G,1966JETPL...4...78Z} at energies $>10^{19}\,$eV by slowing-interactions of cosmic ray photons with the microwave background radiation, which has been observationally confirmed \citep{2008PhRvL.101f1101A}.

We conclude that high energy neutrons are energy inefficient for interstellar communication.

\section{Gravitational waves}
\label{gravitational_waves}

\subsection{Production and transmission}
Gravitational waves (GW) are produced by asymmetric acceleration. In a binary system, the GW frequency is twice the orbital frequency, and the amplitude is determined by the change of the mass distribution. It is trivially possible to produce (low amplitude) GWs, in fact any movement of our bodies emits gravitons isotropically. It is however difficult to form a beam with this radiation, and to produce high amplitude GWs.

To our knowledge, the idea to use GWs for communication was first substantiated by \citet{1977PoAn...10...39P} and \citet{1980toky.iafcQ....S} who describe the need to focus the waves, and the authors suggest to use the ``gravitational fields of massive objects with spherical symmetry'', i.e. lensing.

The advantage of GW signals is the complete transparency of the entire universe within out cosmological event horizon, similar to Neutrinos. 

While the velocity of gravitational waves was historically debated to be superluminal \citep{1998PhLA..250....1V,2001astro.ph..6350D}, recent observations of a binary neutron star coalescence by LIGO show coincident signals from the gravitational waves and gamma-rays within $\sim1.7\,$s, indicating that the velocity of GWs is equal, or very close to, the speed of light \citep{2017ApJ...848L..12A}.

\subsection{Beacons}
GW detectors observe the intensity, or amplitude, which falls off as $1/d$ with distance $d$ whereas the flux of an isotropic electromagnetic source drops as $1/d^2$. Without the penalty of the inverse square law, gravitational waves have an advantage for observation (and thus communication) over large distances. They are preferable beacons for cases where the free-space loss of photons exceeds the efficiency penalty (or difficulty) of GW production. For sources (transmitters, t) of the same power, $P_{\rm t, \gamma}=P_{\rm t, GW}$, the receiver (r) flux ratio is $P_{\rm r, \gamma} / P_{\rm r, GW} = 1/d$. For identical receiver sensitivity, photons are preferable as long as $P_{\rm t, \gamma} > d P_{\rm t, GW}$. In other words, at $100\times$ greater distance, the photon power needs to be $100\times$ larger to compensate for its free space loss. 

A BH merger emits $10^{47}$\,J as gravitational waves, as observed by LIGO \citep{2016PhRvL.116f1102A}. For comparison, a type Ia supernova has an energy release of $10^{42}$\,J in photons and $10^{40}$\,J as GWs \citep{2016MNRAS.461.3296N}. Therefore, in this category of most powerful beacons, GW are superior in detectability over Mpc and Gpc distances. The advantage is reduced because of increased technical difficulty for GW detection.

There are considerable issues with the artificial creation of high-amplitude high-frequency GWs. The large masses and velocity changes require large energies for the acceleration and deceleration of the bodies. Signals can only be injected through such changes, and the data rate will be slow because of mass inertia. Such a communication scheme appears energetically wasteful.

\subsection{Lensing}
Gravitational lensing occurs in the same way for GWs as it does for photons. An important difference is that the commonly used geometrical optics approximation holds only as long as the wavelength is much smaller than the Schwarzschild radius of the lens mass, so that diffraction is small \citep{2010PhRvL.105y1101S}. This is the case for IR photons where $\lambda=1\,\mu$m and $r_g = 2GM_{\odot} /c^2 \approx 2,950$\,m is the Schwarzschild radius of the sun. However, GWs with $\lambda < r_g$ have frequencies $>10^5$\,Hz, so that artificial GWs for lensing need to have $>10^5$\,Hz. 

Astrophysical GW sources are BH mergers with frequencies of $1 \dots 10^3$\,Hz, binaries with $10^{-4} \dots 1$\,Hz, and a small stochastic background of lower ($10^{-10} \dots 10^{-6}$\,Hz) frequencies. No natural sources are expected to exist for frequencies $>10^3$\,Hz.

\subsubsection{Backward foci}
Lensing is typically described from the observational side, where the distance between the lens and the observer is typically larger, or of the same order, compared to the distance between the source and the lens. 

Gravitational lensing of GWs increases the energy flux by a magnification factor $\mu > 0$, and the strain amplitude is amplified by $\sqrt{\mu}$ \citep{2017PhRvD..95d4011D}. The gain has a maximum on the axis \citep{1975Ap&SS..34L...7B}
\begin{eqnarray}
\label{eq_3}
\mu_{max}=4\pi\frac{r_g}{\lambda} \approx 12.57
\end{eqnarray}

for $\lambda \approx r_g$ which is low compared to short wavelength lensing of particles \citep[e.g., $10^9$ for IR photons,][]{2017arXiv170605570H}. 

More suitable lenses would be more massive. At the upper end, the center of our galaxy is the supermassive black hole, Sgr~A* \citep{2002Natur.419..694S} with a mass of $M_{\rm bh}=4.02\pm0.2\times10^6\,M_{\odot}$ at a distance from earth of $R_{\rm bh}=7.86\pm0.18$\,kpc \citep{2016ApJ...830...17B}. We can calculate its Schwarzschild radius as $r_{\rm g}=2\,GM_{\rm bh}/c^2\approx1.21\times10^{10}\,{\rm m}\approx17.1\,R_{\odot}\approx0.08\,{\rm au}$, and the apparent size as seen from earth as $\theta_{\rm bh}=3600\times180\times2\,r_{\rm g} / \pi R_{\rm bh} \approx 20.2\,\mu$as. A detector close the the BH has a useful gain of $10^4$ for frequencies of $10^5$\,Hz following eq.~\ref{eq_3}.

\subsubsection{Forward foci}
The maximum BH forward gain for an isotropic radiator occurs if it is placed near the event horizon, where a substantial fraction of the flux is focused. Such a forward lensing has been suggested in the literature, ``A star can produce a forward point focus of extreme magnification. A Schwarzschild black hole has an infinity of forward and backward foci, where there are two types of forward line foci and one kind of backward conical foci.'' \citep{1991LNP...390..299V}. Using the forward focus (for IR lasers) was also suggested by \citet{Jackson2015}.

In this scenario, the distance between the source and the lens is much smaller than the distance between the lens and the observer. The problem for an object near the Schwarzschild radius is the high gravity of $F=GM/r^2\approx 4 \times 10^6$\,gee, implausibly high for advanced games of cosmic snooker producing GWs \citep{2010Natur.466..406B}. After all, the objects need to remain stationary (and can not be in orbit) to keep the alignment between source, lens and receiver.

Gravity decreases to 1\,gee at a distance of $621\,r_{\rm g}=49.3\,{\rm au}$. At this distance, only a small part of isotropic flux ends up in the Einstein ring, making the forward focus of the BH unattractive.

\subsection{Encoding and noise}
In the framework of quantum field theory, the graviton is a hypothetical elementary particle that mediates the force of gravitation. Usable dimensions are frequency and amplitude of the GW through the graviton spin; no polarization or charge is expected. 

With astrophysical sources typically occurring at frequencies from $10^{-3}\dots10^3$\,Hz, no natural sources are reasonably observable through lensing which starts at $>10^5$\,Hz.

\subsection{Assessment}
GWs are energetically inefficient for communication, except perhaps over the very largest distances. Strong GW lensing requires high masses $>10^6\,M_{\odot}$.

\section{Neutrinos}
\label{neutrinos}
The author Stanislaw Lem envisioned Neutrino communication in his 1968 masterpiece ``His Master's Voice'' \citep{lem1999his}. It was first mentioned in the scientific literature, in passing, by \citet{1972Sci...177..163A}. In the same decade, it was discussed extensively \citep{1977Sci...198..295S,1977PoAn...10...39P,1979CosSe...1....2P}, but was long perceived as ``so difficult that an advanced civilization may purposely choose such a system in order to find and communicate only with ETCs at their own level of development.'' \citep{1979AcAau...6..213S}. Interstellar usage scenarios include clock synchronization \citep{1994QJRAS..35..321L}, directed beam communication \citep{2008AcPPB..39.2943S,2009PhLB..671...15L} and exotic scenarios such as using neutrinos to modify the periods of Cepheid variable stars as Morse-code like beacons \citep{2008arXiv0809.0339L,2010NuPhA.844..248P,2012ConPh..53..113L,2015ApJ...798...42H}.

\subsection{Production}
On earth, neutrinos are artificially produced in fission reactors, nuclear bombs, and in accelerators.

\subsubsection{Fission reactors}
In reactors, about 4.5\% of fission energy is radiated away as antineutrino radiation with a peak (maximum) energy around 4\,MeV (10\,MeV). For interstellar communication purposes, this flux is sufficiently high (45 MW out of a GW reactor), but the emission is wide-angle \citep[of order steradian,][]{2000PhRvD..61a2001A}, impossible to modulate on short (sub-second) timescales, and the low energy makes detection difficult (section~\ref{sub:neutrino_detection}).

\subsubsection{Nuclear bombs (and supernovae)}
Nuclear bombs produce antineutrinos from the fission process, and both neutrinos and antineutrinos in case of a fusion stage. These isotropic flashes occur within a short time; 99.99\% of the energy is released in $8\times10^{-8}$\,s. A large (100 MT) fusion bomb has an energy release of $4\times10^{17}$\,J (or 4\,kg mass equivalent), of which $\approx5$\% or $2\times10^{16}\,{\rm J}\approx10^{38}\,{\rm eV}$ is in neutrinos. With a spectral peak of $\approx10$\,MeV, this translates into a flash of $10^{31}$ particles. With isotropic radiation, the flux at a distance to the nearest star (1.3\,pc) is $6\times10^{-10}$\,m$^{-2}$. Due to the low cross-section of MeV neutrinos, a planet-size detector would be required to detect this flash. To make a more easily detectable neutrino flash, the flux would need to be much larger. Supernovae emit of order $10^{46}$\,J ($10^{29}$\,kg mass equivalent) in neutrinos, a number sufficiently large to be detectable over kpc distance with small detectors \citep{1987Natur.326..135B,2009APh....31..163P}. For comparison, the energy release of SNe is much smaller in photons ($10^{42}$\,J) and GWs ($10^{40}$\,J) \citep{2016MNRAS.461.3296N}.

\subsubsection{Accelerators}
The first accelerator-based neutrino beam used a proton beam hitting a beryllium target, producing pions, which decayed into GeV neutrinos \citep{1962PhRvL...9...36D}. This principle is still used in modern accelerators. 

The ultimate (and not yet built) neutrino beam would be a ``Neutrino Factory'', generating neutrinos by the decay of muons stored in a particle accelerator. Muons decay after $2.2\,\mu$s (ms at GeV energies) into a muon neutrino and an electron anti-neutrino.  In this short time, the muons must be made, collimated, and accelerated; a very challenging technical problem \citep{2013arXiv1308.0494D,2015arXiv150201647D}. These concepts are large (km size) machines with low ($\ll 1\%$) efficiency, comparable in immaturity to the first lasers in the 1960s. Physical efficiency limits are speculated to be of order 0.1\% to 10\% \citep{2009PhLB..671...15L,2010PhLB..692..268H}.

\subsection{Extinction and detection}
\label{sub:neutrino_detection}
The minimal cross-section of neutrinos \citep{2012RvMP...84.1307F} is both a blessing and a curse. On the one hand, interstellar extinction is negligible in all circumstances. On the other hand, for low-energy (MeV) neutrinos, the mean free path is more than a light year of lead, requiring implausibly large detectors if a relevant fraction of particles shall be retrieved. The cross-section peaks for energies near 6.3\,PeV due to the \citet{1960PhRv..118..316G} resonance.  At this energy, the detection fraction in $1\,{\rm km}^3$ of water is 1\%.

\subsection{Communication demonstration}
Real-world neutrino communication has been demonstrated recently at the Fermilab through 240\,m of solid rock, at a data rate of 0.1\,bits/s \citep{2012MPLA...2750077S}. The neutrino beam at the Main Injector (NuMI) is one of the most intense neutrino beams worldwide, producing an arcmin beam peaking near 3.2\,GeV of mostly muon neutrinos, with a wall-plug power of 400\,kW \citep{2005physics...8001K}. The small distance between transmitter and receiver allowed for negligible free-space loss due to beam angle widening. Still, the low detector efficiency (section~\ref{sub:neutrino_detection}) results in a low data rate, showing that significant improvements in neutrino beams and detectors are required for a real world application.

The technology is potentially useful for submarine communication \citep{2010PhLB..692..268H}. To establish even a low data rate communication between base and submarine, $10^{14}$ muon neutrinos per second at 150 GeV are required. Even with a theoretical transmitter efficiency of 10\%, a wall-plug power of 65\,MW would be required.

\subsection{Focusing}
\label{neutrino_focusing}
An ideal neutrino producing accelerator is based on muons, and thus its beam width scales as in section~\ref{acc}:

\begin{equation}
\theta_{\rm beam} = \frac{1}{\gamma} \approx \frac{10^{-4}}{E_{\mu} {\rm [TeV]}}
\end{equation}

where $\gamma$ is the relativistic boost factor of a muon, and $E_{\mu}$ is its energy \citep{2009JInst...4T5001I}. The beam angle at GeV (TeV, PeV) energies is $6^{\circ}$ (21\,arcsec, 21\,mas). For PeV Neutrinos, we can approximate the capacity (in units of bits per energy) as

\begin{equation}
\label{eq_neutrino}
C_{\nu} \approx
10^{-10}
\left(\frac{d}{1\,{\rm pc}} \right)^{-2}
\left(\frac{V}{100\,{\rm km}^3} \right)
\,\,(\rm {bits\,J^{-1}})
\end{equation}

where $d$ is the distance
and $V$ is the water/ice detector volume.
Estimates valid for all energies can not be given in closed form, because the cross section of Neutrinos is a complicated function of energy \citep{2012RvMP...84.1307F}. For example, Neutrinos at GeV instead of PeV energy require a collector volume larger by three orders of magnitude, in order to achieve the same collection fraction.

Even large detectors can not compensate for the focusing disadvantage of Neutrinos. For comparison, the opening angle of diffraction-limited optics is $\theta_{\rm optics} = 1.22 \lambda / D_{\rm t}$. For $D_{\rm t}=1$\,m, $\theta_{\rm beam} = \theta_{\rm optics} $ at $\lambda=82$\,nm (15 eV), a difference of $7\times10^{10}$ in energy for the same beam width. In other words, focusing neutrinos into a beam requires $7\times10^{10}$ more energy in a particle accelerator compared to a meter-sized mirror (using photons).

\subsection{Capacity and encoding}
The capacity for neutrino communication (in bits per neutrino) can be calculated in the same way as for other particles. The usable dimensions are time of arrival, energy, and particle/antiparticle; there is no polarization. Instead, neutrinos come in three flavors (electron, muon, tau), but these cannot be used for encoding due to oscillations \citep{1998PhRvL..81.1562F}. The probability of neutrino oscillations follows a function of the ratio $L/E$ where $L$ is the distance traveled and $E$ is the energy. The distance-energy for a probability of order unity is $\approx1,000\,{\rm km\,GeV}^{-1}$, so that even for high-energy neutrinos (6.3\,PeV), oscillations randomize the flavors for $L>42$\,au, approximately the distance to Pluto.

\subsection{Gravitational lensing}
Gravitational lensing of neutrinos follows the same laws as for photons, where the bending angle is inversely proportional to the impact parameter $b>R_{\odot}$ of a light ray with respect to the lens. This inverse bending angle is called astigmatism and produces a caustic focal line.

\subsubsection{Focal length}
The difference to photons is that neutrinos (and GWs) also pass \textit{through} the sun, resulting in a shorter minimum focal length of $z_{\rm 0,\nu}=23.5\pm0.1$\,au, about the distance of Uranus \citep{2000PhRvD..61h3001D,2008ApJ...685.1297P}, compared to $z_{\rm 0,\gamma}=R_{\odot}^2/2\,r_g \approx 546$\,au for photons, where $r_g = 2\,GM_{\odot} /c^2 \approx 2,950$\,m is the Schwarzschild radius of the sun \citep{1964PhRv..133..835L,1979Sci...205.1133E}.

\subsubsection{Aperture size}
Classical lensing collects the photons from the very thin Einstein ring surrounding the star, whose width is equal to the receiver size, and whose circumference is a circle with an impact parameter $b=\sqrt{z/z_{\rm 0}}$ where $z$ is the heliocentric distance. Then, the area of collected light is $A_{\gamma}=\pi ((b + w) ^2 - b^2)$. For a meter-sized detector in the lens plane, the corresponding classical aperture is 74.6\,km \citep{2017arXiv170605570H,2017arXiv170305783T}.

In contrast, a transparent gravitational lens has an effective aperture of a fraction the stellar size, as most of the rays pass through the star. Based on solar density gradient models, different estimates of the effective lens radius have been calculated, ranging from $R_{\rm lens}=0.024\,R_{\odot}\approx16,700\,{\rm km}$ \citep{2008ApJ...685.1297P} to $R_{\rm lens}=0.17\,R_{\odot}\approx118,300\,{\rm km}$ \citep{2000PhRvD..61h3001D,2005ApJ...628.1081N}. In realistic cases, $A_{\nu} \gg A_{\gamma}$. However, the Neutrino aperture is fixed and independent of the detector size, in contrast to the photon case.

\subsubsection{Size of the caustic}
The caustic focal line of classical lensing extends unbroadened towards infinity. Transparent lensing, in contrast, produces a focal point, followed by an extending cone \citep{1981ApJ...244L...1B,2008ApJ...685.1297P}. The minimum point spread function width in the focal plane (at a distance of 23.5\,au) is dominated by irregularities inside the sun, namely convection cells (contributing less than 50\,m), oblateness (1\,m), and spots (few m). The total effect has been estimated to about $1 \dots 50$\,m \citep{2000PhRvD..61h3001D} and more detailed modeling seems necessary before the deployment of a real detector. Of course no model is perfect, and the use of a real neutrino lens detector would inversely teach us a lot about the sun's interior; perhaps worth a mission on its own.

\subsubsection{Detector mass equivalent}
For a lens radius of $R_{\rm lens}=16,700 \dots 118,300$\,km which gets compressed in the image plane to a minimum (diameter) of $R_{\rm image}=0.025 \dots 0.005$\,km, we can calculate the corresponding mass gain as $R_{\rm lens}^2 / R_{\rm image}^2 \approx 5\times10^{11} \dots 6\times10^{16}$. Therefore, placing a receiver mass of one ton in the image plane will yield as many neutrino detections as having $10^{11} \dots 10^{16}$ tons in the receiver on earth.

\subsubsection{Transmission}
The transmission probability for neutrinos of different energies and flavors through the SGL has been studied in detail by \citet{1999hep.ph...10510E}. There is essentially a cutoff near 100\,GeV, so that higher energy neutrinos are absorbed inside the sun (and re-radiated into arbitrary directions), while those of lower energy pass through almost unaffected.

\subsubsection{Station keeping}
To observe a single source continuously, the detector (mass) must be kept stationary along the axis of source, sun, and detector. Therefore, the detector can not rotate around the sun, and must instead counter its (small) gravitational pull $F$ to keep in place,

\begin{equation}
F=\frac{G M_{\odot} m}{z_{\rm 0}^2}\approx0.01\,{\rm N.}
\end{equation}

for $m=1,000$\,kg and $z_{\rm 0}=23.5$\,au. This value is small as even the smallest rocket engines have kN thrust, so that station keeping appears trivial. The location must be kept within meter accuracy, due to the small size of the caustic.

\subsubsection{Resolution}
The magnification of the SGL is very high and can be calculated from geometrical optics. The image is smaller than the object by $R/z$, with $R$ as the distance to the object. For $z=23.5$\,au and our closest neighbor Alpha Centauri, $R=1.3$\,pc, we get $R/z=10^4$. Thus, an earth-sized (12,756\,km) source at this distance would appear with a size of 1\,km in the image plane. A meter-sized telescope in the image plane would resolve an area of 10\,km$^2$. Consequently, nearby astrophysical sources such as stars appear spatially resolved (and only part of the flux is collected). For imaging of extended objects, a scanning flight would be required. For interstellar communication purposes, transmitter apertures must be smaller than $D_{\rm r}R/z$. With such a high resolution, only one source can be observed at a time. To target a new object, the detector must be slewed.

\subsubsection{Point spread function}
The relation between the relativistic momentum and the wavelength $\lambda$ is defined by the Broglie equation, $\lambda=hc/E$ where $h$ is Planck's constant. For example, a GeV neutrino will have a wavelength of $10^{-15}$\,m, much ($10^5\times$) smaller than the size of an atom. In the image plane of the solar gravitational lens, the point spread function (PSF) width for $\lambda=10^{-15}$\,m is \citep[][their Eq. 142]{2017arXiv170406824T}

\begin{equation}
\rho\approx 4.5~\Big(\frac{\lambda}{1\,\mu{\rm m}}\Big)\frac{b}{R_\odot}~{\rm cm} \approx 10^{-11}\,{\rm m}
\end{equation}

which is much smaller than the image spread caused by the lens imperfections and can therefore be neglected.

\subsection{Assessment}
Neutrinos are not ideal as beacons. Even large fusion bombs are insufficient to be detectable from the nearest stellar systems due to the low cross-section of the emitted MeV neutrinos. The larger fluxes of Supernovae are detectable; and one might imagine mid-sized fusion events (e.g. a moon-mass, $10^{22}$\,kg) to be detectable over kpc distance, but this appears extremely wasteful.

Gravitational lensing is attractive if the placement of large masses ($>1,000$\,kg) in the outer solar system is cheap. Precisely, the mass gain is $10^{11}$ to $10^{16}$ which can be used to calculate cost efficiency compared to a large planet-based neutrino detector, e.g. using water or ice. A disadvantage for a lens collector is the directivity. The resolution is so high that only one distant object can be observed at a time, while a classical (e.g. cubic, spherical) detector is sensitive to incoming neutrinos from all directions.

Regarding directed energy: The need for large accelerators, large receivers and low efficiency keeps neutrinos in the field of a ``difficult'' technology. Speculative future advances, such as (femto-)technology to manipulate nuclear matter \citep{bolonkin2009femtotechnology}, may allow for more efficient neutrino capture, and thus higher detection fractions. It is presently unclear if there is a physically plausible solution to get in the 50\% (or even 1\%) efficiency range available for photons. Neutrino communication also suffers from the focusing efficiency issue: their beam width is $10^{10}$ times wider as for photons of the same energy. Finally, the large detectors disqualify neutrinos for the use on board of small probes.

\section{Relaxing constraints}
\label{irrational_aliens}
Our initial assumptions of what ET values might be incorrect. We had assumed that ET favours more information over less, fast over slow, has energy limits, and wants to build less machinery rather than more, all else equal. We will now relax these constraints one by one and re-evaluate our analysis.

\subsection{Little information is enough}
Although humans are a curious species, an advanced civilization might be bored of factual communication with others. To exchange just a few bits (``I'm here!''), a beacon is sufficient. Isotropic microwave beacons are the most energy-efficient direct emitters, but are still very costly if they shall be seen over large distances \citep{2010AsBio..10..491B,2010AsBio..10..475B}. Modulating a pulsator can be much easier at the same visibility (section~\ref{arti}). In this scenario, we should focus on a deep all-sky survey at GHz frequencies, as well as look for strange stellar-like object, such as Boyajian's star \citep{2016MNRAS.457.3988B,2016ApJ...829L...3W}. With sufficiently good instruments, the need for active beacons vanishes. Smaller and smaller macroengineering objects become remotely visible, such as ``Clarke Exobelts''\citep{2018ApJ...855..110S}, industrial air pollution \citep{2014ApJ...792L...7L}, climate change, or catastrophic nuclear wars \citep{2016IJAsB..15..333S}. If such signals are detectable as by-catches of regular observations, the cost of (beacon style) communication is zero, making it the most efficient approach.

\subsection{Preference of complicated technology}
Perhaps as an entry-barrier for lesser civilizations, ET might favor more complicated technology. 

If this is to be done at the cost of higher energy usage, or at the cost of complicated (more expensive) machinery, an obvious choice would be $\gamma$-rays, Neutrinos or massive particles such as Neutrons.

At the cost of slower communication (time of arrival), one would send inscribed matter in the form of probes, or perhaps bullets, which need to be found and/or decelerated by the target civilization. Depending on how small such an object is, we would find it inside our solar system only at advanced stages of exploration.

\subsection{Relaxed energy limits}
If energy efficiency is not considered important, wider beams are an obvious choice, as they have the advantage of more civilizations being potentially located inside the cone. In this lighthouse scenario, low photon energy communication, i.e. microwaves, are clearly favored because they maximize the information per unit area. Tight beams with wasteful energy usage would use heavy particles such as Neutrons.

\subsection{Unimportant Time delay}
One could argue that entities in advanced civilizations live much longer, e.g. through hibernation, biological modifications, mind-uploading into interchangeable substrates such as silicon, or other trans-humanist mechanisms. 

If time delay is not important, the obvious choice would be to send inscribed matter. This can be the most energy-efficient choice \citep{2004Natur.431...47R,2017arXiv171210262H}, because it can be done at almost arbitrarily low velocities ($\approx 20$\,km\,s$^{-1}$), and thus low energies. Also, an artifact can arrive at the destination in total, in contrast to a beam which is wider than the receiver in all realistic cases.

\subsection{Major gaps in our knowledge of physics}
\label{we_know_nothing}
Perhaps there exist yet unknown particles whose spec sheets exceed that of photons. While we can not know their nature, we can state the advantages we require them to posses in order to exceed photons. They must be either cheaper (in terms of energy requirements per beam width), faster, more robust, have more encoding modes, simpler to use, or a combination of these.

It appears implausible that yet unknown particles exist which are easier to use than photons or classical matter.

Regarding robustness, there are already known particles which have excellent extinction performance, e.g. high-energy photons or Neutrinos. Axions (section~\ref{hypo}), so they exist, would have zero extinction. However, they are more difficult to receive than photons, because their frequency needs to be known exactly. It also appears that these can only be received mono-chromatically (because the cavity is tuned to only one frequency), reducing the number of modes, and thus the number of bits per photon in any encoding scheme. While the total effect is unclear without knowing their exact properties, axions appear not to be superior over photons, and certainly not superior by more than a factor of a few.

For quicker communication, tachyonic particles would travel faster than light \citep{1967PhRv..159.1089F}, and therefore be preferred over photons, all else equal. However, faster than light signals are generally assumed to violate causality and are therefore unlikely to exist, although hypothetical speculations remain \citep{2011arXiv1112.4714C}.

It is also difficult to imagine particles which are preferred because of a higher number of encoding modes. The number of modes has a logarithmic influence on the data rate and is therefore only of marginal benefit \citep{2017arXiv171205682H,2018arXiv180106218H}.

Furthermore, we can exclude all hypothetical particles with high masses \citep[e.g., the Neutralino with $>300$\,GeV,][]{2010pdmo.book..142E}, because it is too costly in terms of required energy to accelerate these compared to mass-less photons.

Overall, it appears unlikely that any hypothetical particles exceeds keV photon performance by more than a factor of a few. Unless our understanding of physics has fundamental gaps, photons are the rational choice for most communications.

\subsection{Multiple incorrect assumptions}
Combining two or more of these arguments yields arbitrary results. For example, if advanced ET favors energy inefficiency \textit{and} little information, it would be reasonable to create SNe. An approach combining slow data rate \textit{and} expensive machinery could be to build artificial occulters.

\section{Conclusion}
\label{conclusion}
We have benchmarked all known information carriers against photons, including electrons, protons, neutrinos, inscribed matter (probes), gravitational waves and artificial megastructures such as occulters. We compared the speed of exchange, information per energy and machine sizes, lensing performance, and complexity.

Isotropic beacons with low data rates can not be constrained with our current level of understanding, and many options appear similarly attractive.

In point-to-point communications, we assumed that ET favours more information over less, fast over slow, simplicity over complexity, and has energy limits. We explored the consequences when dropping these assumptions one by one. If the assumptions hold, then photons are superior to other carries by orders of magnitude. If speed is not crucial, sending probes with inscribed matter would be preferred. 

\pagebreak
\section{References}
\bibliographystyle{elsarticle-harv}
\bibliography{references_elsevier}
\end{document}